\documentclass[openacc]{rstransa}
\usepackage{graphicx}
\usepackage[outercaption]{sidecap}    
\usepackage{caption}
\usepackage{subcaption}

\definecolor{os}{rgb}{1.0,0.0,0.0} 
\definecolor{bf}{rgb}{0.65,0.16,0.16} 

\begin{document}

\title{Acoustic-gravity wave propagation characteristics in 3D radiation hydrodynamic simulations of the solar atmosphere}

\author{
B. Fleck$^{1}$, M. Carlsson$^{2,3}$, E.~Khomenko$^{4,5}$, M. Rempel$^{6}$, O.~Steiner$^{7,8}$ and G. Vigeesh$^{7}$}

\address{$^{1}$ESA Science and Operations Department, c/o NASA/GSFC Code 671, Greenbelt, MD 20771, USA\\
$^{2}$Rosseland Centre for Solar Physics, University of Oslo, Postboks 1029 Blindern, N-0315 Oslo, Norway\\
$^{3}$Institute of Theoretical Astrophysics, University of Oslo, Postboks 1029 Blindern, N-0315 Oslo, Norway\\
$^{4}$Instituto de Astrof{\'i}sica de Canarias, E-38205 La Laguna, Tenerife, Spain\\
$^{5}$Departamento de Astrof{\'i}sica, Universidad de La Laguna, E-38205, La Laguna, Tenerife, Spain\\
$^{6}$High Altitude Observatory, NCAR, P.O. Box 3000, Boulder, Colorado 80307, USA\\
$^{7}$Leibniz-Institut f{\"u}r Sonnenphysik (KIS), Sch{\"o}neckstrasse 6, 79104 Freiburg, Germany\\
$^{8}$Istituto Ricerche Solari Locarno (IRSOL), Via Patocchi 57, 6605 Locarno-Monti, Switzerland
}

\subject{\textcolor{bf}{Astrophysics, Solar System, Space Exploration, Sun}}

\keywords{sun, atmosphere, waves, simulations}

\corres{Bernhard Fleck\\
\email{bfleck@esa.nascom.nasa.gov}}

\begin{abstract}
There has been tremendous progress in the degree of realism of three-dimensional radiation magneto-hydrodynamic simulations of the solar atmosphere in the past decades. Four of the most frequently used numerical codes are Bifrost, CO5BOLD, MANCHA3D, and MURaM. Here we test and compare the wave propagation characteristics in model runs from these four codes by measuring the dispersion relation of acoustic-gravity waves at various heights. We find considerable differences between the various models. The height dependence of wave power, in particular of high-frequency waves, varies by up to two orders of magnitude between the models, and the phase difference spectra of several models show unexpected features, including $\pm180^\circ$ phase jumps.
\end{abstract}

\maketitle



\section{Introduction}
Driven by the tremendous increases in computational resources over the past decades, computational astrophysics has become an important and rapidly growing discipline of astronomy, including solar physics. Radiation hydrodynamic simulations [1] of the solar atmosphere have come a long way since the pioneering work by {\AA}.\,Nordlund [2]. While even today's most sophisticated models are numerical experiments that should not be trusted as accurate renderings of the Sun, recent three-dimensional radiation hydrodynamic simulations have given fascinating new insights into numerous physical phenomena occurring in the Sun's atmosphere, as e.g., sunspots [3, 4, 5], solar surface convection [6, 7], the generation of spicules [8], coronal heating [9, 10], solar chemical composition [11,12], or flares in the corona [13], to name but a few. 

Four of the most frequently used numerical codes are Bifrost [14, 15], CO5BOLD [16], MANCHA3D [17, 18, 19], and MURaM [20, 21]. Some of these models were benchmarked for their average properties in the near-surface layers, by comparing their average stratifications as well as their temporal and spatial fluctuations (e.g., the root mean square (RMS) of granular contrast and vertical velocities) [22]. The objective of the present study is to investigate ("benchmark") the wave propagation characteristics and damping of acoustic-gravity waves in selected model runs from these four codes by measuring the dispersion relation of acoustic-gravity waves at various heights. Once we have a good understanding and validation of the simulations, we will address science questions such as the height dependence of the energy flux of acoustic-gravity waves [23]  and the propagation characteristics in the chromosphere (what is the cause of the very high phase speeds measured in the chromosphere? [24]).

\section{Method}
How do we test the various models for their wave propagation characteristics? We do so by measuring the dispersion relation of acoustic-gravity waves $k_z = k_z(\omega, k_h, c_s, g, \tau_R)$, with $\omega = 2\pi/\nu$, $\nu$ being the frequency of the wave, $k_h = 2\pi/\lambda$ the horizontal wavenumber, $c_s$ the speed of sound, $g$ the surface gravity of the Sun, and $\tau_R$ the radiation damping time. For a wave with velocity amplitude $v ~ e^{i(\omega t - k_z z)}$, the phase difference between the velocity signal at heights $z_1$ and $z_2$ is $\Delta \phi_{21} = k_z(z_2 - z_1)$, 
i.e., is a direct measure of $k_z$, if the height difference $(z_2 - z_1)$ is known.

\begin{figure}[!tbp]
\centering
\begin{minipage}[b]{0.45\textwidth}
\includegraphics[width=\textwidth]{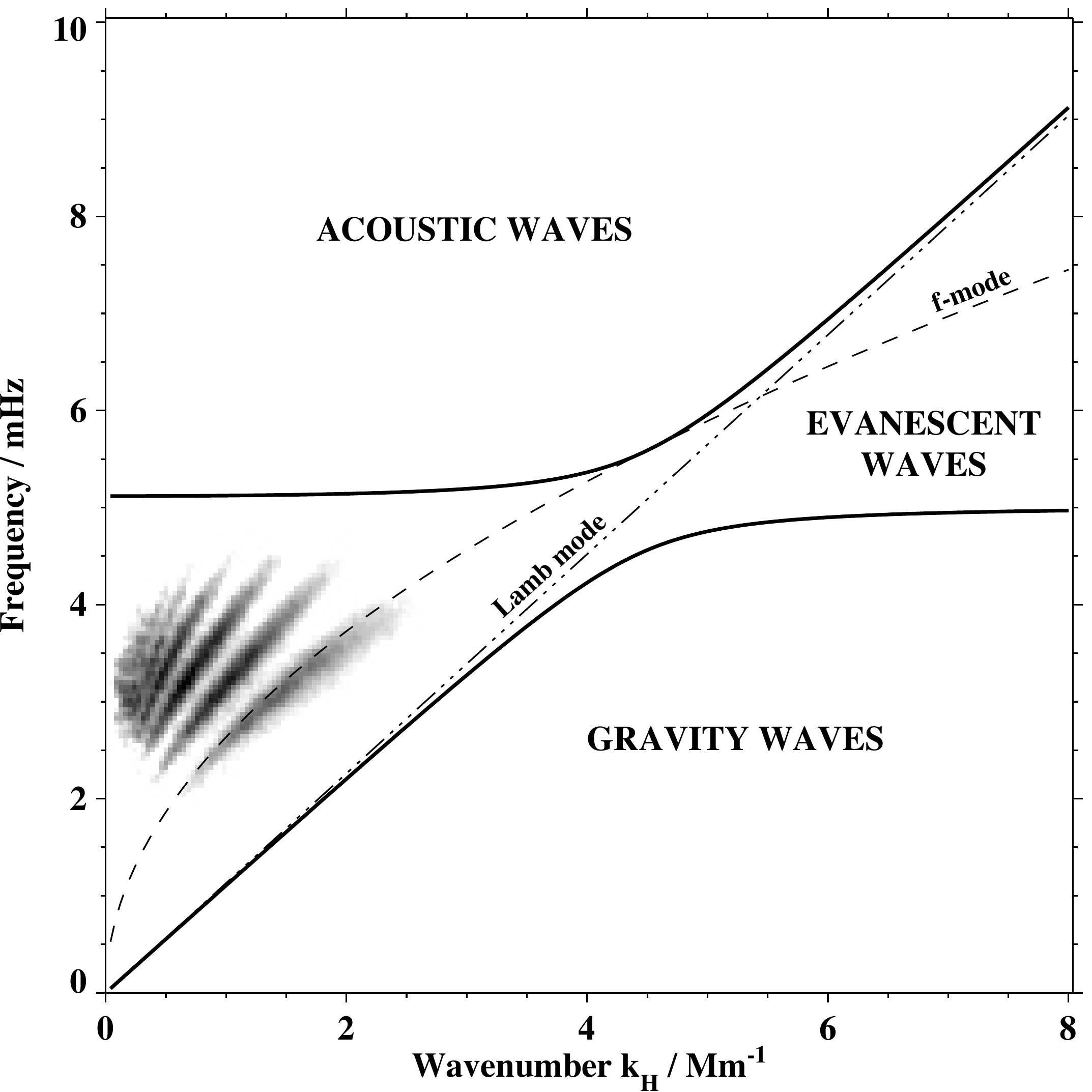}
\caption{Diagnostic diagram of an isothermal atmosphere with $c_s=7.1$\,km\,s$^{-1}$, $\gamma=5/3$,
and $g=274$\,m\,s$^{-2}$.} 
\label{fig:diagnostic-diagram}
\end{minipage}
\hfill
\begin{minipage}[b]{0.5\textwidth}
\includegraphics[width=\textwidth]{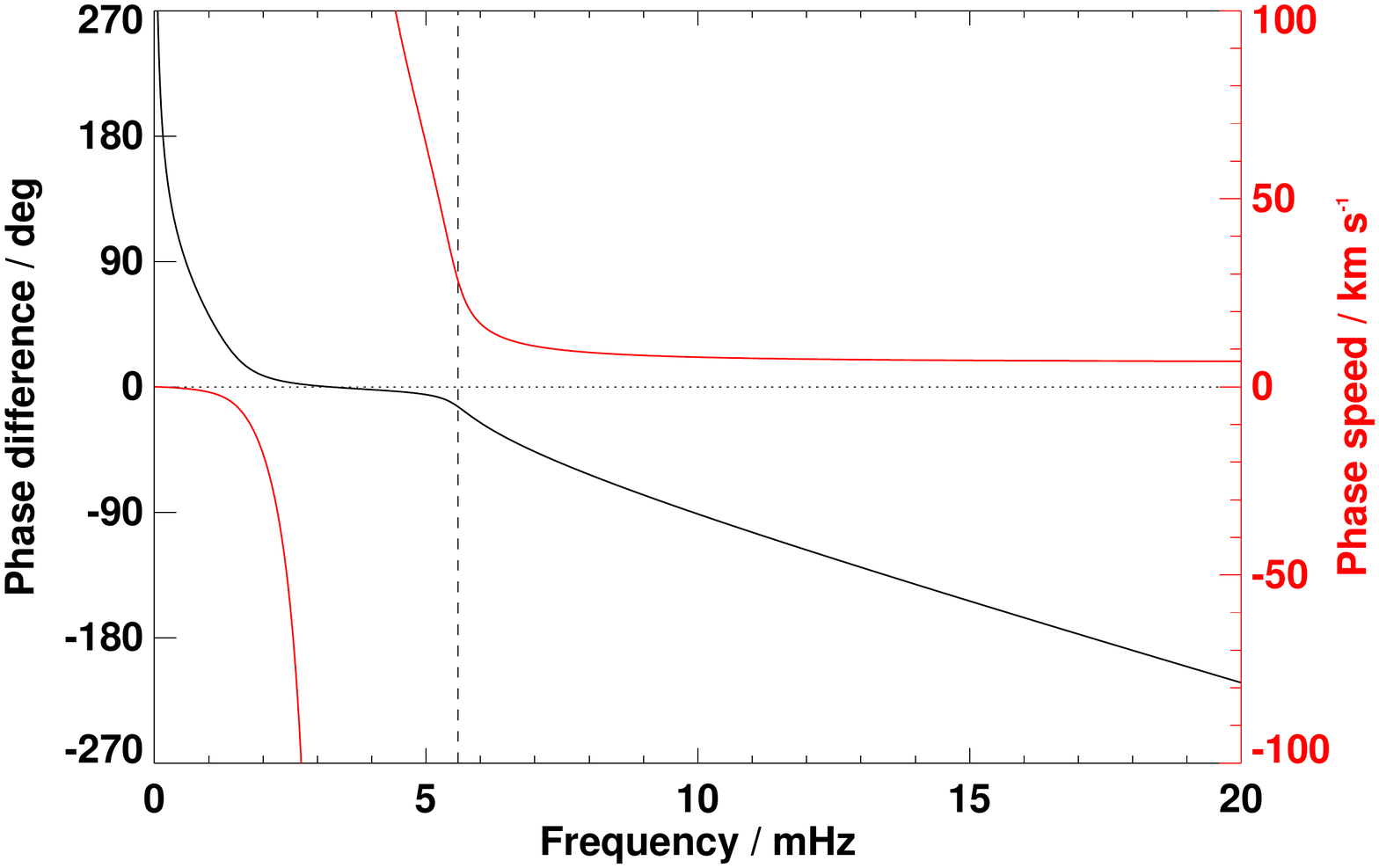}
\caption{Black line: Phase difference spectrum of acoustic gravity waves between two layers separated by $\Delta z$=200\,km in an isothermal atmosphere with $c_s$=6.5\,km\,s$^{-1}$, $\tau_R =180$\,s,  and $k_h=1.5$\,Mm$^{-1}$.  Red lines: Phase speed in km\,s$^{-1}$. The dashed line marks the acoustic cutoff frequency, $\nu_{\rm ac}=5.6$\,mHz.}
\label{fig:souffrin_model}
\end{minipage}
\end{figure}

For an isothermal, stratified atmosphere with constant radiative damping the dispersion relation of acoustic-gravity waves can be expressed analytically [25] (see also 
eqs.\,(3)--(6) of [26]). Important frequencies in this context are the acoustic cutoff frequency $\omega_{ac} = 2\pi \nu_{ac} = {\gamma g} / {2 c_s}$ and the Brunt-V{\"a}is{\"a}l{\"a} frequency $\omega_{BV} = \sqrt{(\gamma -1)}{g/c_s}$. To remind the reader of the frequency-wavenumber ranges where acoustic (and gravity) waves propagate, we display a diagnostic diagram of an isothermal atmosphere with $c_s = 7.1$\,km\,s$^{-1}$, $\gamma=5/3$, $g=274$\,m\,s$^{-2}$ in 
Fig.\,\ref{fig:diagnostic-diagram}. The solid lines mark the locations where $k_z = 0$. Waves can only propagate in the acoustic and gravity wave branches where $k_z$ is real. In between these two branches lies the region of evanescent waves, where $k_z$ is imaginary and the waves are exponentially damped. Over-plotted in Fig.\,\ref{fig:diagnostic-diagram} is the location of the f-mode ($\omega = \sqrt{g k_h}$, dashed line), the Lamb mode of horizontally propagating waves ($\omega = c_s k_h$, dash-dotted line), and the location of the 5-min p-mode ridges displayed on an inverse B/W color scale. A theoretical phase difference spectrum for an isothermal atmosphere with sound speed 6.5\,km s$^{-1}$ following [25] is shown in Fig.\,\ref{fig:souffrin_model} in black, and the resulting phase speed $c_{\rm ph} = \Delta z / \Delta \phi * \nu$ in red.  The vertical dashed line marks the acoustic cutoff  frequency (5.6\,mHz).


In this paper we use the sign convention to subtract the phase of the higher layer from the phase of lower layer, i.e. upward propagating waves show up with a negative phase difference. One can easily discern three regimes in Fig.\,\ref{fig:souffrin_model}: (1) a transition to a linear decrease of the phase difference at frequencies above the cutoff frequency, the signature of running acoustic waves, with the phase speed approaching the speed of sound; (2) a range of very small phase differences between $\approx 2$ and 5\,mHz (evanescent waves), with exceedingly high phase velocities, and (3) at very low frequencies a steep rise to large positive phase differences, the signature of gravity waves, which due to their transversal nature show up with downward propagating (positive) phase differences for upward propagating energy.

\begin{figure}
\centering
\includegraphics[width=0.8\textwidth]{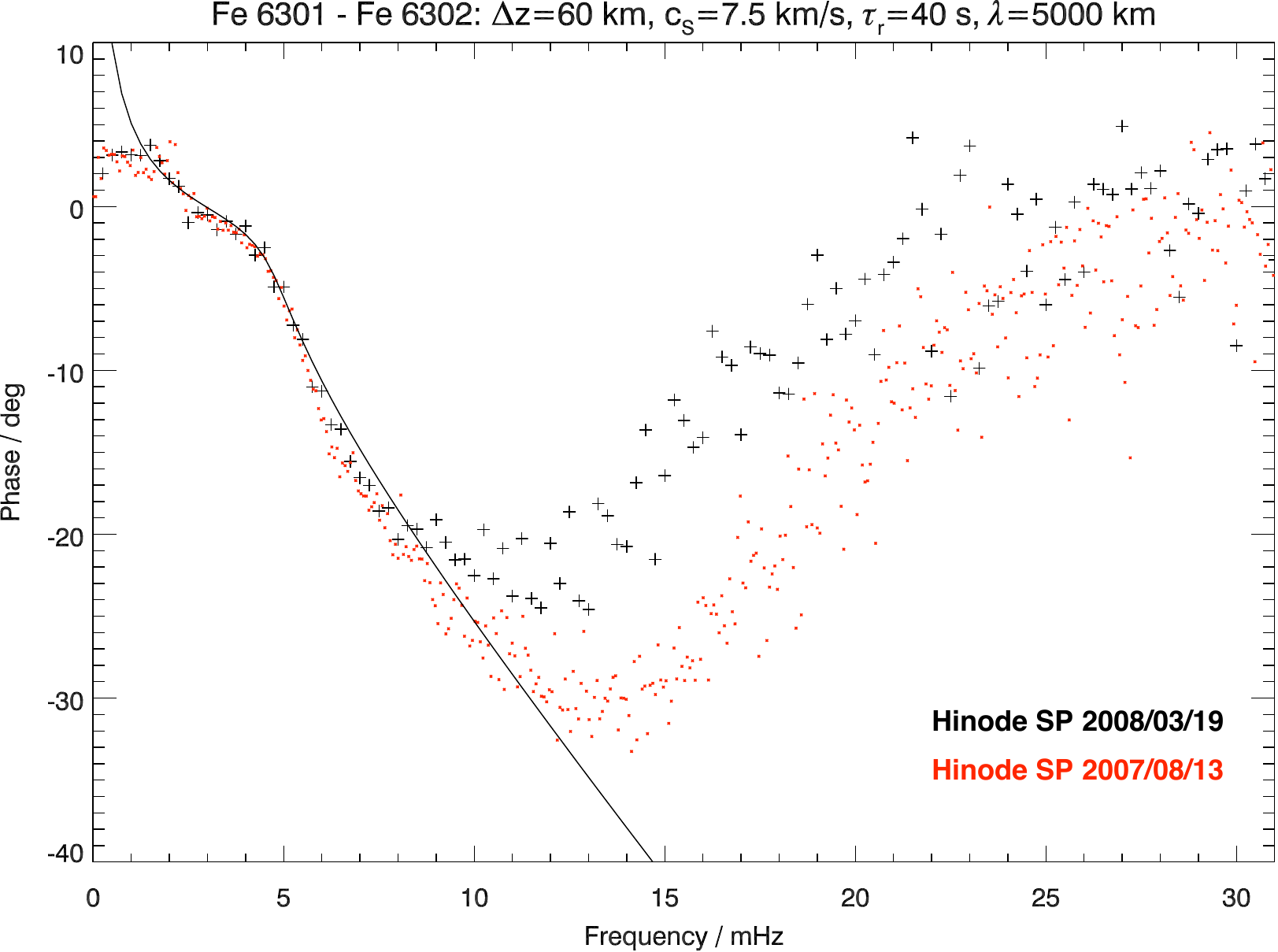}
\caption{Observed phase difference spectra between Fe 6302 and Fe 6301 derived from two Hinode SP sit-and-stare runs in the quiet Sun. The parameters used for the theoretical curve (solid line) are $\Delta z=60$\,km, $c_s=7.5$\,km\,s$^{-1}$, $\tau_R=40$\,s, and $k_h= 1.25$\,Mm$^{-1}$.}
\label{fig:Hinode-SP-phase-diff}
\end{figure}

As can be seen in Fig.\,\ref{fig:Hinode-SP-phase-diff}, this rather simple model fits observed phase difference spectra remarkably well (see also e.g. Fig.\,1 of [24] or Fig.\,1 of [26]). Fig.\,\ref{fig:Hinode-SP-phase-diff} shows the phase difference between the two Fe lines at 6302 and 6301 \AA, obtained from two sit-and-stare runs with the Hinode Spectropolarimeter (SP, [27]) on 2007/08/13 and 2008/03/19. The two Fe lines are formed in the middle photosphere in the range of  250 to 320\,km [28] with a difference in formation height of $\Delta z \approx 60$\,km. The cycle time of both series was 16\,s. The 2007/08/13 series (red dots) had a duration of 237\,min and was taken with the long slit (1024 pixels) at full spatial resolution (0.16\,arcsec/pixel).  The 2008/03/19 series (black crosses) had a shorter duration (66\,min) and was taken with a shorter slit (384 pixels) in binned mode (0.32 arcsec/pixel). As can be seen,  the higher signal to noise ratio of the longer and higher resolved series taken in 2007 allows to follow the linear decrease of the phase difference expected for running acoustic waves up to about 12\,mHz ($\approx$ 80\,s period). At higher frequencies, the coherence between the two signals is too low to retrieve reliable phase differences, which pull back to near zero (the phase of the dominant signal near 3\,mHz). We should emphasize that, from an observational point of view, 12\,mHz is pretty much the upper limit for which reliable and meaningful phase difference spectra can be determined from photospheric lines, and some of the effects we will discuss below occur at frequencies significantly higher than that, i.e., in a range where we have no constraints from observations.

\section{Models}

Below we introduce the models we used for this investigation, in alphabetical order.
We notice that these models are quite different with respect to spatial extension and resolution, duration, magnetic fields, and physics included so that caution is indicated when making comparisons.

\subsection{Bifrost [14, 15]}
We used various quiet Sun ("qs" with "salt and pepper" (sap) magnetic field) model runs, with different sizes and different lower boundary conditions. The Bifrost models are the only ones with a realistic chromosphere, transition region, and corona. At the bottom boundary there is a pressure node, and the top boundary of the Bifrost simulations is transmitting through characteristic boundary conditions. More important for wave reflections than an imperfect upper boundary  are (physical) reflections from strong gradients in the wave speeds in the transition region and from wave conversion at inclined fields.

\subsubsection{qs024031\_sap}
\begin{itemize}
\item $768\times768\times768$ 
computational cells with $\Delta x,y = 31$\,km, $\Delta z $ variable,  $\Delta z \approx 12$\,km in the photosphere
\item 968 frames at $\Delta t = 10$\,s cadence (164\,min duration)
\item $23.8 \times 23.8 \times 16.8$\,Mm$^{3}$
\item Lower boundary at $-2.5$\,Mm, upper boundary at 14.3\,Mm
\item $<B_z>=0$\,G, $<|B|> = 44$\,G and $B_{\rm RMS} = 78$\,G at $z=0$\,km
\end{itemize}

\subsubsection{qs006023\_t100}
\begin{itemize}
\item $256\times 256\times 512$ computational cells, with $\Delta x,y = 23$\,km, $\Delta z $ variable,  $\Delta z \approx14$\,km in the photosphere
\item 306 frames at $\Delta t = 10$\,s cadence (51\,min duration)
\item $5.9 \times 5.9 \times 10.5$\,Mm$^{3}$
\item Lower boundary at $-2.5$\,Mm, upper boundary at 8\,Mm
\item Timescale for lower boundary: 100\,s
\item $<B_z>=-2.5$\,G, $<|B|> = 11$\,G and $B_{\rm RMS} = 47$\,G at $z=0$\,km
\end{itemize}

\subsubsection{qs006023\_t007}
\begin{itemize}
\item Same as qs006023\_t100, but with a different lower boundary condition and only 154 frames (25\, min duration)
\item Timescale for lower boundary: 7\,s
\item $<B_z>=-2.5$\,G, $<|B|>= 8-20$\,G and $B_{\rm RMS} = 54$\,G at $z=0$\,km
\end{itemize}

\subsubsection{qs012023\_t100}
\begin{itemize}
\item Same as qs006023\_t100, but $2 \times$ FOV ($2 \times$ the number of computational cells in both x and y) and only 242 frames (40\, min duration)
\item $11.8 \times 11.8 \times 10.5$\,Mm$^{3}$
\item Timescale for lower boundary: 100\,s
\item $<B_z>=-2.5$\,G, $<|B|> = 11-24$\,G and $B_{\rm RMS} = 59$\,G at $z=0$\,km
\end{itemize}

The last three models have not reached saturation of the small scale dynamo and the average magnetic filed is still increasing with time. The range above is from the first to the last snapshot, the RMS for the last snapshot. (The value given for model qs006023\_t007 is for snapshot number 657, which is further in time than the snapshots available when we performed this analysis.) 

\subsection{CO5BOLD [16]}
\subsubsection{d3\_cpchange0p3  and d3\_cpchange1p0}
The CO5BOLD models used here are non-magnetic models computed by using the MHD module but setting the initial magnetic flux density to zero. The two models (cp0p3 and cp1p0) differ by their lower boundary condition. Model cp1p0 has a stiffer lower boundary than cp0p3 in the sense that the gas pressure fluctuation in the bottom cells is damped more quickly and kept more tightly to the mean pressure at the bottom boundary. We analyzed both types of model runs with grey and non-grey radiation transfer (RT). The lower boundary is open in the sense that the fluid can freely flow in and out of the computational domain under the condition of vanishing total mass flux. 
The upper boundary is constructed to be transmitting so that acoustic waves can leave the computational domain with little reflection at the boundary. This is achieved with stress-free conditions for the velocities, viz., ${\rm d} v_{x,y,z} /{\rm d} z = 0$ and likewise for the internal energy. The density is assumed to decrease exponentially with height into the ghost layers, with a scale height set to a controllable fraction of the local hydrostatic pressure scale height. The layers of ghost cells are located outside the computational domain proper.
\begin{itemize}
\item 720 frames at 10 s cadence for grey RT runs (120\,min duration)
\item 240 frames at 30 s cadence for non-grey RT runs (120\,min duration)
\item $480 \times 480 \times 120$  computational cells with $\Delta x,y = 80$\,km, $\Delta z $ variable in the convection zone,  $\Delta z=20$\,km in atmosphere ($\tau > 1$)
\item $38.4 \times 38.4 \times 2.8$\,Mm$^3$
\item Lower boundary: $\approx -1.5$\,Mm
\item Upper boundary: $\approx 1.3$\,Mm
\end{itemize}
These models are special for CO5BOLD simulation runs because of the rather unfavorably large aspect ratio of their computational cells of $\Delta x,y/\Delta z = 4$ in the atmosphere. They correspond to model \texttt{d3t57g45v50gv} of [29], which reports on the slowly growing $p$-mode amplitude of this model. 

\subsection{MANCHA3D [17, 18, 19]}
The MANCHA3D model used here is from a local dynamo simulation with a mean magnetic field in the range of typical quiet Sun values [30, 31]: $<B_z>=0$\,G, $<|B_z|>  \approx 67$\,G, $<|B|> \approx 110$\,G and $B_{\rm RMS} \approx 140$\,G at $z=0$\,km. While these fields should not have a significant influence on the dynamics, the wave speeds may be modified slightly and there might be additional dissipation. The most significant difference between the MANCHA3D model and the other models lies in the fact that the MANCHA3D model includes ambipolar diffusion, which causes extra dissipation of high frequency waves ($\nu > 10$\,mHz) in the upper photosphere and chromosphere above 0.5\,Mm [17]. The MANCHA3D model studied here uses a open bottom boundary condition with mass and entropy controls that ensure that the model produces the correct solar radiative flux. The top boundary is closed for mass flows, with symmetric boundary conditions (zero gradient) for internal energy and density. The temperature is computed through the equation of state (EOS).

\subsubsection{c3d\_alta10\_battery\_ambi\_noBhyp\_rk4\_10sA\_32bits}
\begin{itemize}
\item 1486 frames at 10\,s cadence (247\,min duration)
\item $288 \times 288 \times 168$ 
computational cells
with $\Delta x,y = 20$\,km, $\Delta z = 14$\,km
\item $5.8 \times 5.8 \times 2.4$\,Mm$^{3}$
\item grey RT
\item Lower boundary: --0.952\,Mm
\item Upper boundary: 1.386\,Mm
\end{itemize}

\subsection{MURaM [20, 21]}
The MURaM models used here are from small-scale dynamo simulations with magnetic fields similar to those in the MANCHA3D simulation. At the lower boundary, the mean pressure and specific entropy in the upflow regions are fixed (to a value that gives the correct solar energy flux). Pressure perturbations are damped (this was needed for stability, but it also means that p-modes are at least partially reflected). Both mass flux and magnetic field are mirrored across the boundary (symmetric boundary condition), i.e., the code allows for both vertical and horizontal flows and magnetic fields. The top boundary is half open, i.e., open for upflows and closed for downflows. This reduces reflections when strong shocks hit the boundary.
\subsubsection{dyn\_6x4x6Mm\_(non\_)grey\_tvd\_2(0)}
\begin{itemize}
\item 3 models in total: 1 grey RT and 1 non-grey RT with normal diffusivity ($\ldots$ tvd\_2), and one model with non-grey RT but much higher numerical diffusivity ($\ldots$ tvd\_0)
\item tvd\_2: $<B_z>=0$\,G, $<|B_z|> \approx 80$\,G,
$<|B|> \approx 150$\,G and $B_{\rm RMS} \approx 225$\,G at $z=0$,\,km
\item tvd\_0: $<B_z>=0$\,G, $<|B_z|> \approx 54$\,G,
$<|B|> \approx 107$\,G and $B_{\rm RMS} \approx 166$\,G at $z=0$,\,km
\item 1800 frames at 2.025\,s cadence (62\,min duration)
\item $384 \times 384\times 256$ 
computational cells
with $\Delta x,y = 16$\,km, $\Delta z = 16$\,km
\item $6.1 \times 6.1 \times 4.1$\,Mm$^{3}$
\item Lower boundary: --2.35\,Mm
\item Upper boundary: 1.74\,Mm
\item Also one much larger model with $1536 \times 1536 \times 512$ cube ($24.6 \times 24.6 \times 8.2$\,Mm$^3$, but only 106 frames at 18.45\,s cadence (32 min duration)
\end{itemize}

\begin{figure}
\centering
\includegraphics[width=0.8\textwidth]{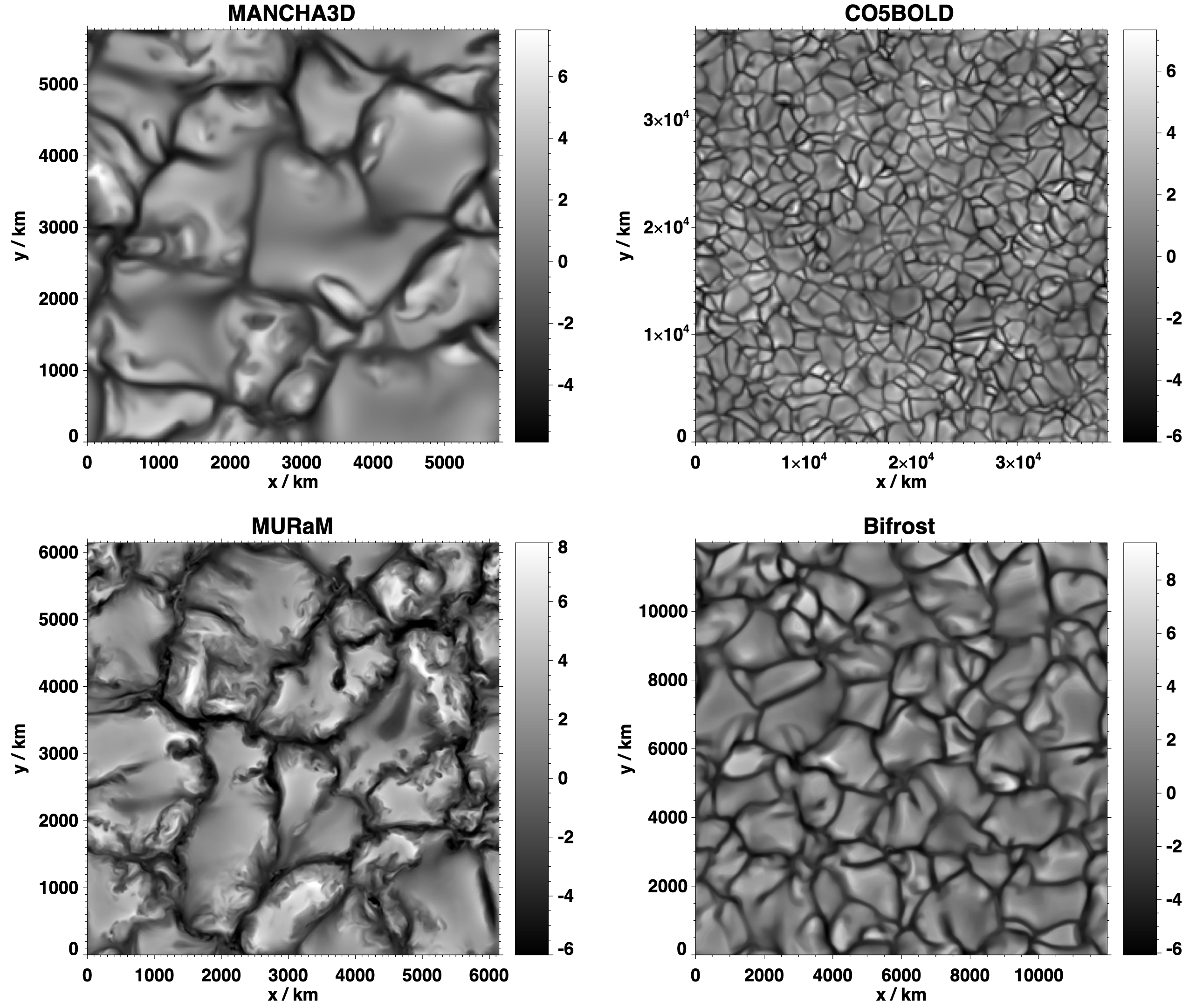}
\caption{Snapshots of the vertical velocity in km s$^{-1}$, positive values are upflows and negative values downflows) at $z=0$\,km in some of the simulations we used, illustrating the largely different fields of view.}
\label{fig:simulations_overview}
\end{figure}

Snapshots of the vertical velocity at $z=0$\,km in some of the models we used are shown in Fig.\,\ref{fig:simulations_overview}, illustrating their considerably different fields of view.

We are confident that wave reflection at the top boundaries is not a significant issue and that the top boundaries are not affecting the results in the domain we are interested in in this study (the photosphere and lower chromosphere). While time distance diagrams (z-t cuts through the simulation cubes) show evidence of downward propagating waves, those appear to originate predominantly in regions of strong gradients in the wave speeds and where shocks form or merge, not at the upper boundaries. This suggests that the upper boundaries are indeed working as designed.  Any residual small amplitude waves that might be caused by imperfections in the upper boundaries will be quickly damped due to the increase in density as they propagate down in the atmosphere.

\section{Results}
\subsection{Power spectra}

\begin{figure}
\centering
\includegraphics[width=1.0\textwidth]{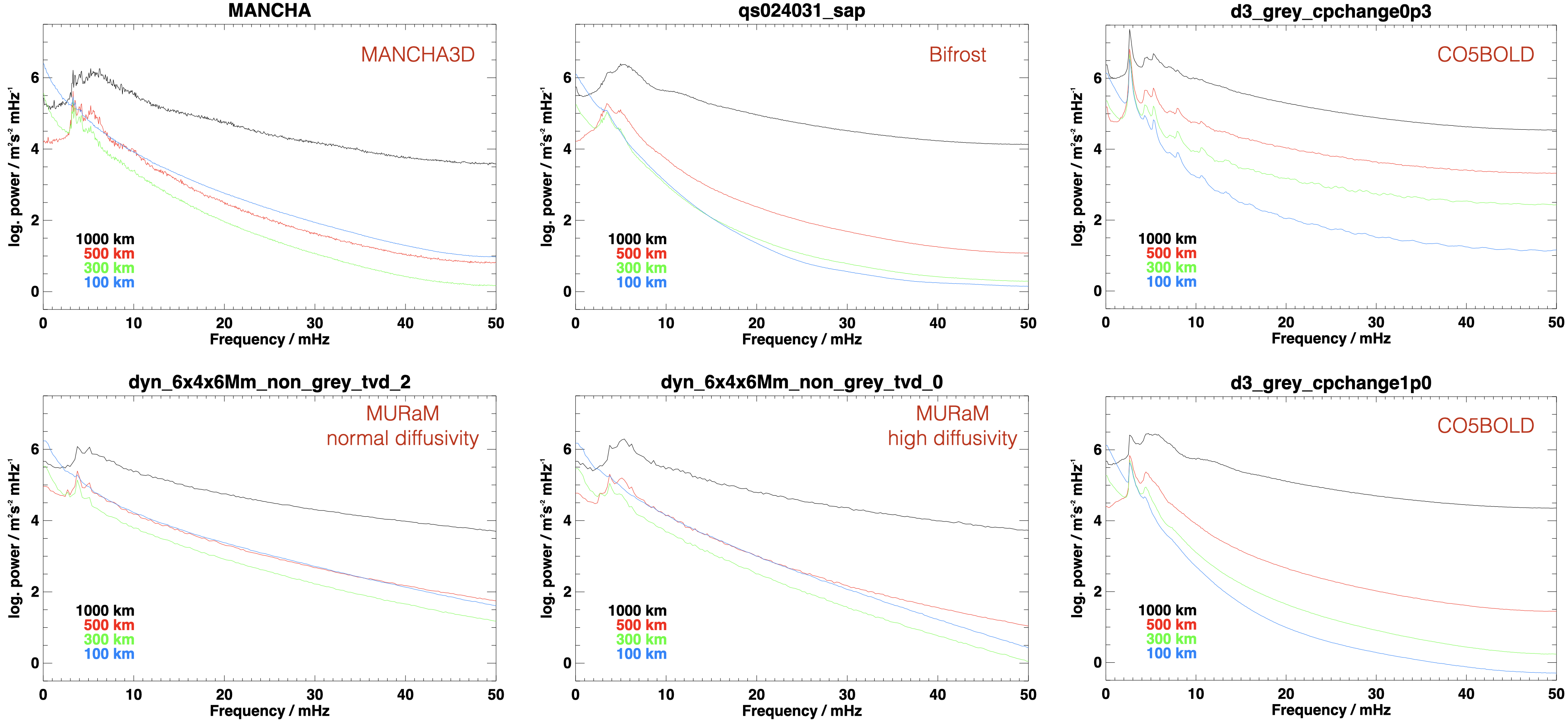}
\caption{Power spectra (logarithmic display) at selected heights (blue: 100 km; green: 300 km; red: 500 km; black: 1000 km), normalized to m${^2}$ s$^{-2}$\,mHz$^{-1}$. Upper left: MANCHA3D; upper middle: Bifrost qs024031\_sap; upper right: CO5BOLD cp0p3; lower right: CO5BOLD cp1p0; lower middle: MURaM high diffusivity; lower left: MURaM normal diffusivity.}
\label{fig:power_spectra}
\end{figure}

\begin{figure}
\centering
\begin{subfigure}{.45\textwidth}
\includegraphics[width=\textwidth]{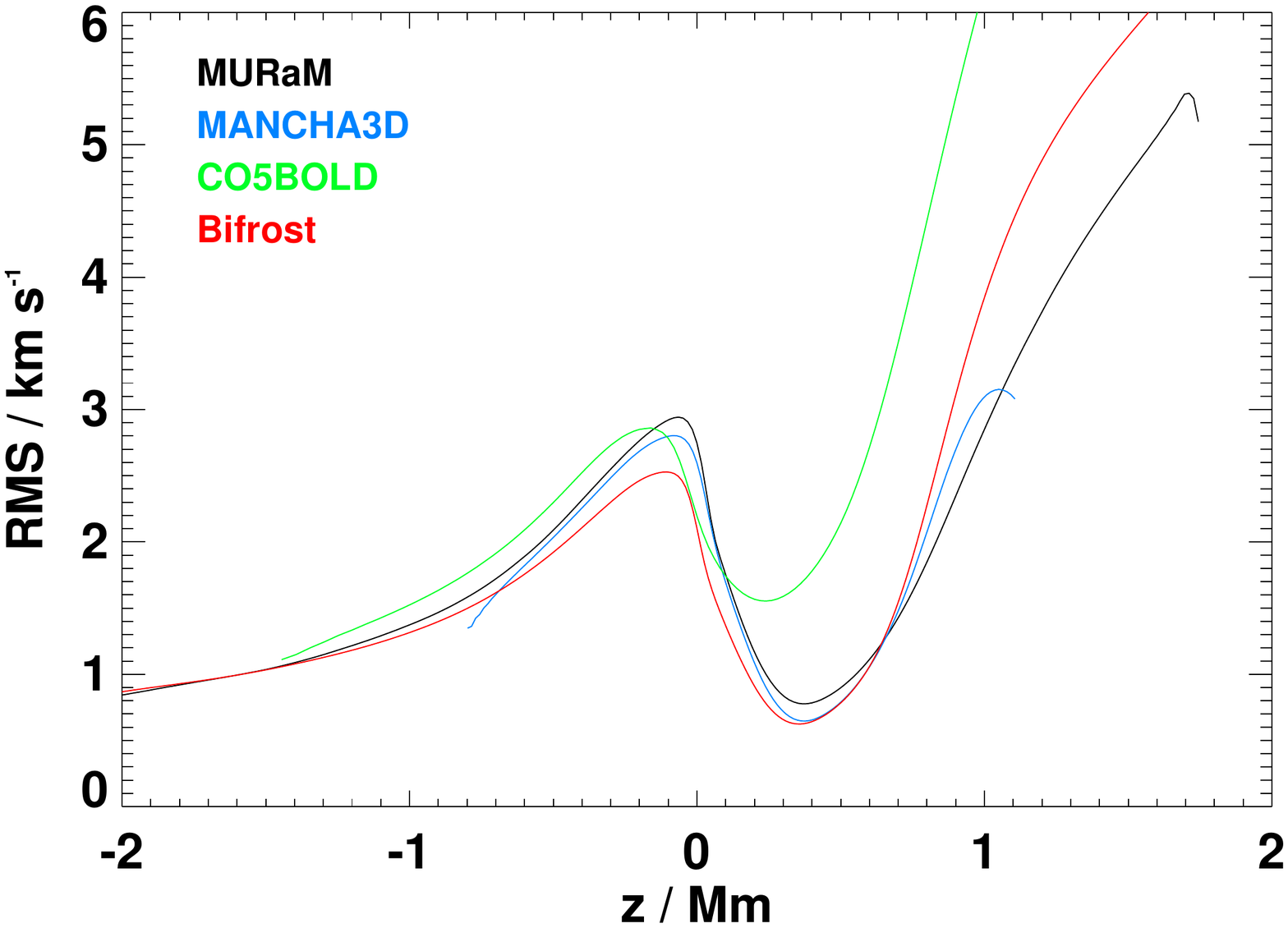}
\end{subfigure}
\begin{subfigure}{.45\textwidth}
\includegraphics[width=\textwidth]{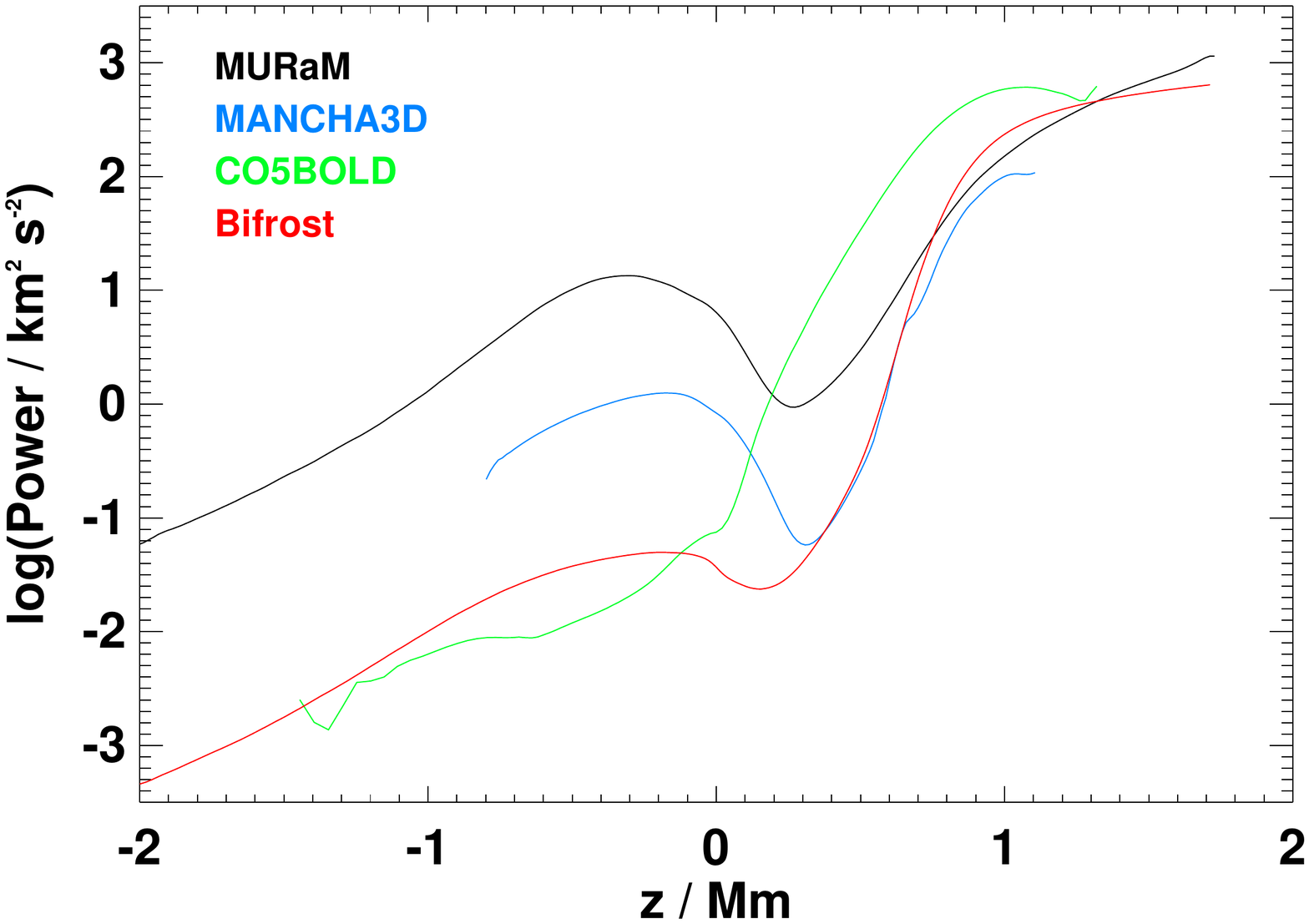}
\end{subfigure}
\caption{Left: Total RMS of the vertical velocity amplitudes vs height in the atmosphere in four selected simulations (see text). Right: Intergrated power in the range 30\,mHz $\le \nu \le 40$\,mHz vs height in these simulations.} 
\label{fig:sim_rms_vs_height}
\end{figure}

In Fig.\,\ref{fig:power_spectra} we display the average power spectra
 $ P(\nu) = \frac{1}{N^2} \sum_{i=1}^{N} \sum_{j=1}^{N} |\hat{v}(x_{i},y_{j},\nu)|^2 $, with  $\hat{v}(x_{i},y_{j},\nu)$ being the Fourier 
 transform
 of the vertical velocity $v(x_{i},y_{j},t)$  at pixel $(x_i,y_j)$, for four selected heights (100, 300, 500, and 1000\,km) 
 for several of the models presented above. While the general appearance is similar, with a peak power near 3 mHz building up in the photosphere and the peak power shifting to near 5 mHz higher up in the atmosphere, the differences of the power evolution with height are considerable. The CO5BOLD 
 models show very prominent power peaks in the 5-min region. These peaks are already visible low in the photosphere in the cross section 
 at 100\,km (the p-modes are even visible in the CO5BOLD snapshot in Fig.\,\ref{fig:simulations_overview}). In the other models the 5-min oscillations become a distinct feature of the power spectra only at larger heights in the middle photosphere (300\,km sections). 
 Noticeable is also the reduced power in the MURaM simulation using a higher diffusivity, in particular at high frequencies in the middle photosphere. Perhaps most striking apart from the strong p-mode resonances in the CO5BOLD simulations are the significant differences of the height-dependence of the high frequency power distributions. Fig.\,\ref{fig:sim_rms_vs_height} illustrates this more clearly. There, we show the RMS of the vertical velocity amplitude of four selected simulations vs height in the solar atmosphere (left figure) and the integrated power in the high frequency range 30\,mHz $\le \nu \le 40$\,mHz vs height (right figure). The models used for this Figure are:  Bifrost qs024031\_sap, the MANCHA3D model described above, the non-grey MURaM run with normal diffusivity, and the grey CO5BOLD cp0p3 run. 
 
All models show a local maximum of the total RMS of the vertical velocity amplitudes just below the bottom of the photosphere (z=0\,km), a rather steep drop by nearly 2\,km s$^{-1}$ to a local minimum in the middle photosphere ($z \approx 400$\,km), followed by a steep rise in the higher layers. We plotted the MANCHA3D model only up to 1100\,km, as the upper 200\,km are affected by a diffusion layer near the upper boundary of the computational box and therefore not valid. As expected from the power spectra in Fig.\,\ref{fig:power_spectra}, the CO5BOLD model 
shows the highest RMS amplitudes.

 While the total RMS of the vertical velocity amplitudes displayed in Fig.\,\ref{fig:sim_rms_vs_height}  shows reasonably good agreement between the models, the power at high frequencies and its height dependence reveals stark differences between the models (right panel 
 of Fig.\,\ref{fig:sim_rms_vs_height}). For instance, the high frequency power in the convection zone of the MURaM simulation is about two orders of magnitude larger than that in the Bifrost simulation. In the latter one, the high frequency power increases very steeply in the photosphere, such that it becomes even bigger than in the MURaM simulation at about 800\,km height. Noteworth is also the fact that the increase of the high frequency power seems to flatten for all models near 1000\,km (the exception being the MURaM simulation, where the flattening is less pronounced). The CO5BOLD simulation is the only one which shows a steady increase of the high frequency power, whereas the other models reveal local minima between 200 and 500\,km.
 
Given the significant differences in the power distribution between the various models, in particular at high frequencies, caution is advised when using them for energy flux studies.

\subsection{Phase difference spectra between selected heights}

We now turn our attention to the phase difference spectra between selected heights in the various models, i.e., the measured dispersion relation between these heights. We selected four heights: 100\,km (lower photosphere), 300\,km (middle photosphere), 500\,km ("temperature minimum" region), and 1000\,km (middle "chromosphere"). We should emphasize, though, that except for the Bifrost models, none of the other models has a proper chromosphere nor a transition region and corona. We display average phase difference spectra as a function of frequency $\nu$, calculated from the average complex crosspower spectra $CP(\nu)$ over the full fields of view: 
\\
\par
$ CP(\nu) = \frac{1}{N^2} \sum_{i=1}^{N} \sum_{j=1}^{N} \hat{v}_{z_2}(x_{i},y_{j},\nu) \hat{v}_{z_1}^*(x_{i},y_{j},\nu)   $  \\
\par
\noindent
with $\hat{v}_{z_2}(x_{i},y_{j},\nu)$ being the Fourier transform of the vertical velocity $v_{z_2}(x_{i},y_{j},t)$  at pixel $(x_i,y_j)$ and height $z_2$, and $\hat{v}_{z_1}(x_{i},y_{j},\nu)$ being the complex conjugate of the Fourier Transform of the vertical velocity $v_{z_1}(x_{i},y_{j},t)$  at pixel $(x_i,y_j)$ and height $z_1$.  The phase difference  $\Delta \phi_{21}$ is the phase of this complex crosspower vector:  
\\
\par
$\Delta \phi_{21}(\nu) = \arctan \frac{\Im{(CP(\nu))}}{\Re(CP(\nu))}$ \\
\par

\begin{figure}
\centering
\includegraphics[width=1.0\textwidth]{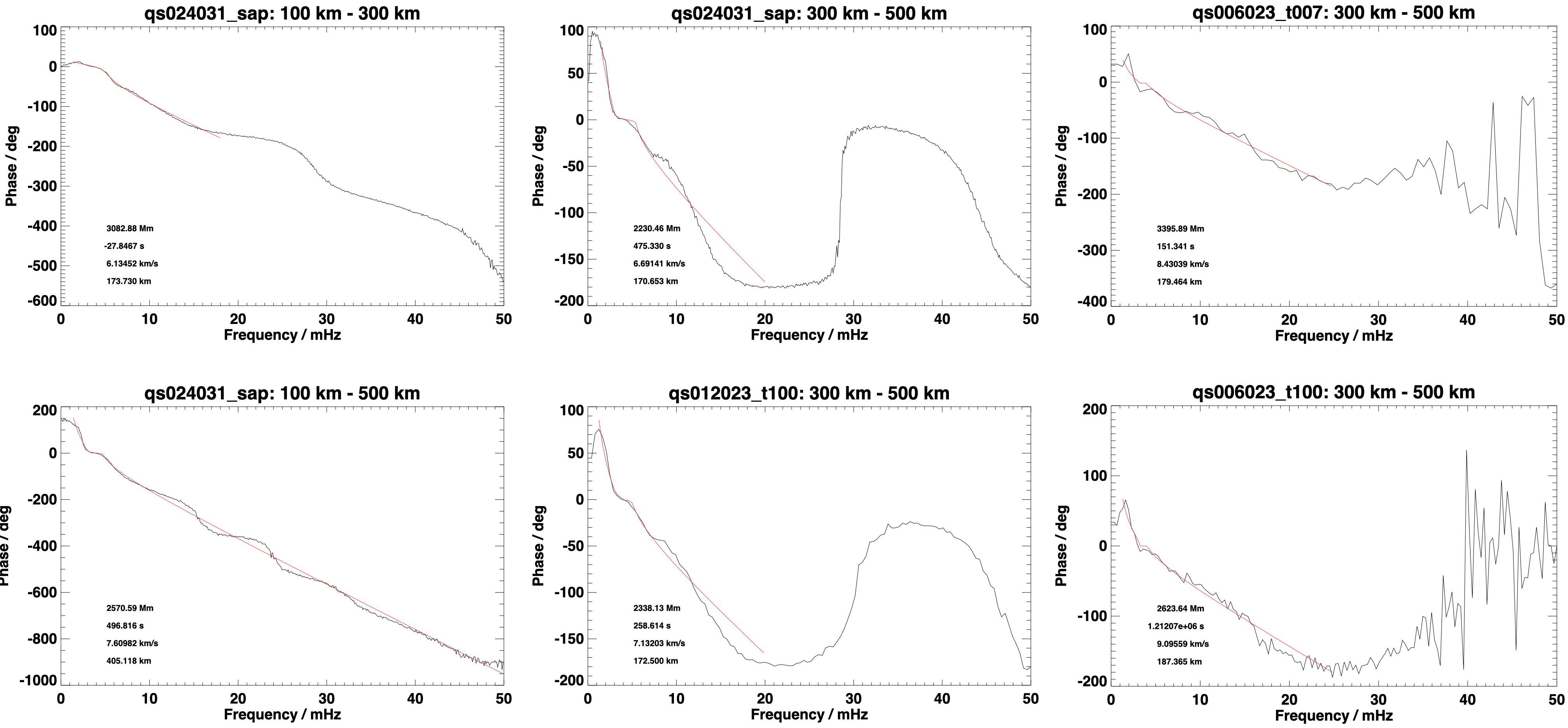}
\caption{Phase difference spectra between selected heights in several Bifrost models (black lines), overlaid by fits using the simple analytical description of acoustic-gravity waves in an isothermal atmosphere by Souffrin [23] (red lines). The paramters of the fits are included in the lower left corners of the diagrams: horizontal wavelength $\lambda = \frac{2 \pi}{k_h}$, radiative damping $\tau_R$, sound speed $c_s$, and height difference between the two layers $\Delta z$.}
\label{fig:1D-phase-Bifrost}
\end{figure}

Fig.\,\ref{fig:1D-phase-Bifrost} shows the phase difference as a function of frequecy for the four Bifrost models for selected height differences (100 km -- 300 km, 100 -- 500 km, and 300 - 500 km). They all show positive values at low frequencies $\nu \lessapprox 3$ mHz, the signature of gravity waves with upward energy flux (see Section 2). Following a plateau in the narrow frequency range between approximately 3 and 5 mhz with very small phase differences (evanescent waves), the phase difference transitions into a linear decrease with frequency, as expected for upward propagating acoustic waves. Here the reader is encouraged to compare the phase difference spectra from the simulations with the expectations from the simple Souffrin model (23) in Fig.\,\ref{fig:souffrin_model} and the observed phase difference spectra in Fig.\,\ref{fig:Hinode-SP-phase-diff}. While there is good qualitative agreement between the phase difference spectra from the simulations and the observations up to about 15 mHz (the upper end of reliable measurements in observations), there are distinct deviations from the simple Souffrin model at higher frequencies. While the Souffrin model predicts a monotone linear decrease with a constant gradient, the phase difference spectra from the Bifrost models show "wobbles", which at certain heights become plateaus (lower left in Fig.\,\ref{fig:1D-phase-Bifrost}), and even full 180$^\circ$ phase jumps (middle column, showing the phase difference between the 300 km and 500 km height). This 180$^\circ$ jump is most prounced in the qs024031\_sap model, which is by far the longest of the Bifrost models we studied, and thus the one with the highest frequency resolution. The 180$^\circ$phase jump is also visible in the qs012023\_t100 model, whereas the qs006023\_t100 models show a different behaviour at $\nu \ge 25$ mHz (but also a clear deviation from the expectations of a monotonous decrease with frequency). 

\begin{figure}
\centering
\includegraphics[width=1.0\textwidth]{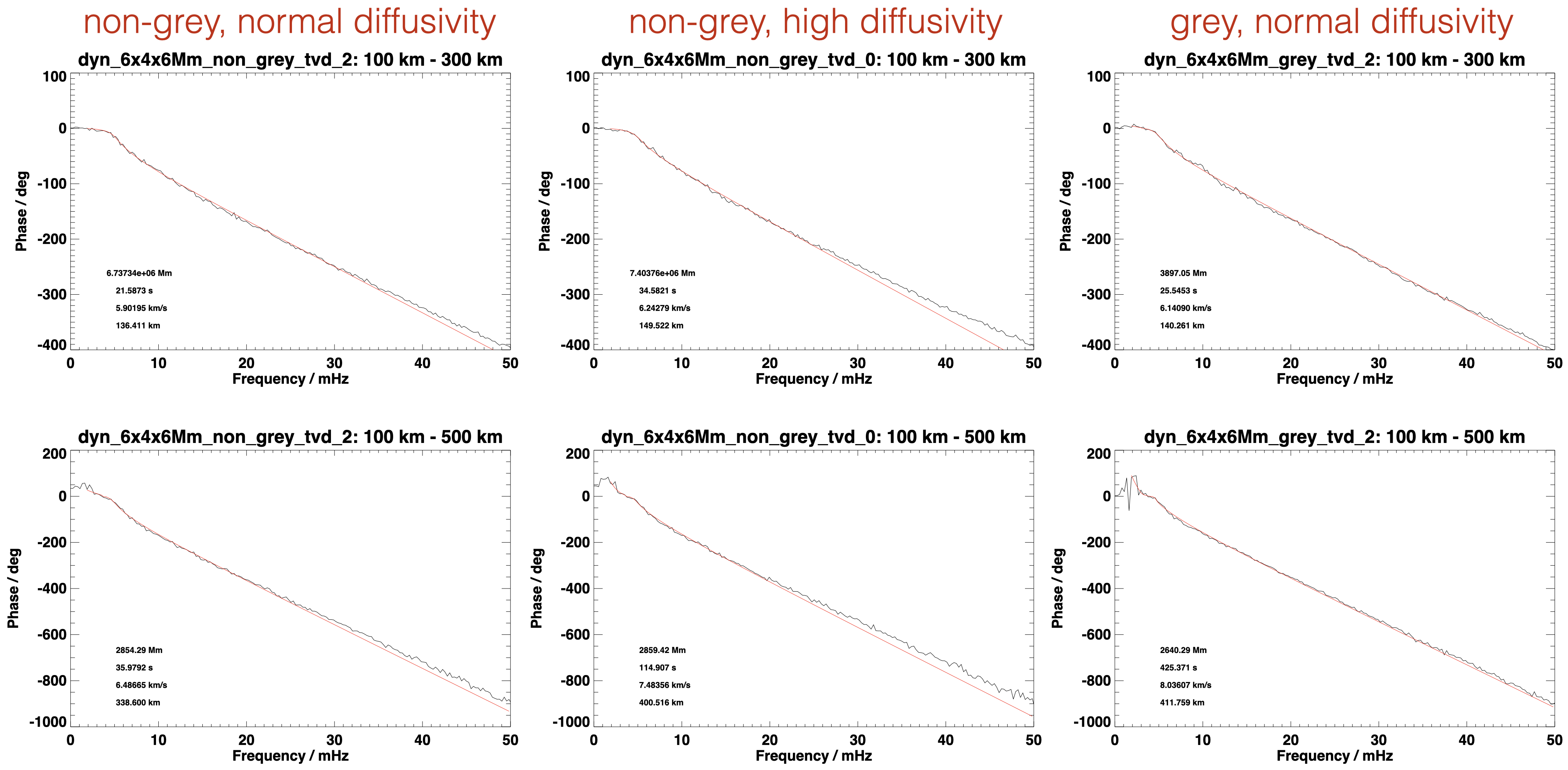}
\caption{Same as Fig.\ref{fig:1D-phase-Bifrost} for the MURaM models.}
\label{fig:1D-phase-Muram}
\end{figure}

\begin{figure}
\centering
\includegraphics[width=1.0\textwidth]{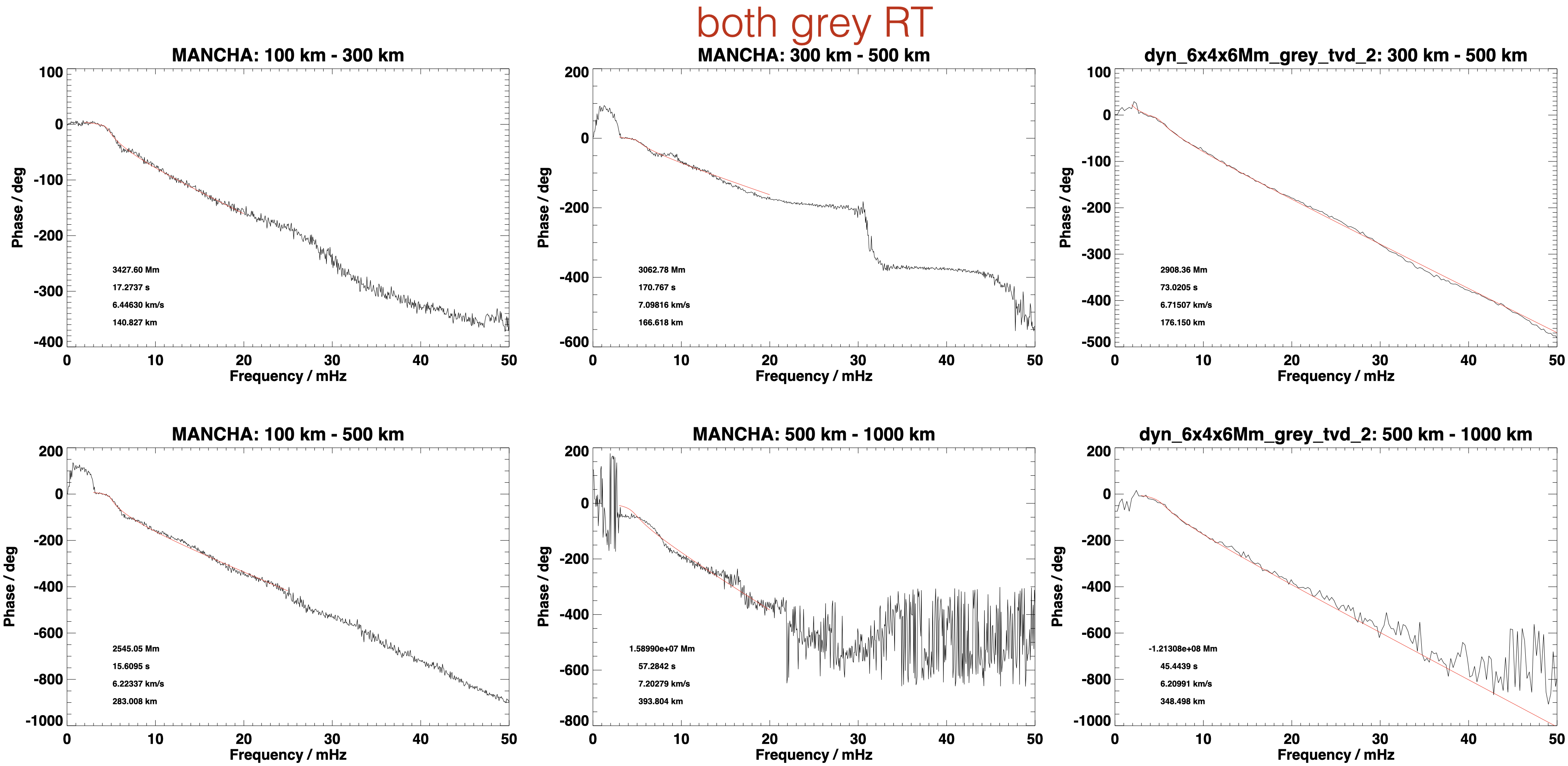}
\caption{Same as Fig.\ref{fig:1D-phase-Bifrost} for the MANCHA3D model (left and middle column), in comparison with a MURaM model (right column)}
\label{fig:1D-phase-Mancha-Muram}
\end{figure}

\begin{figure}
\centering
\includegraphics[width=1.0\textwidth]{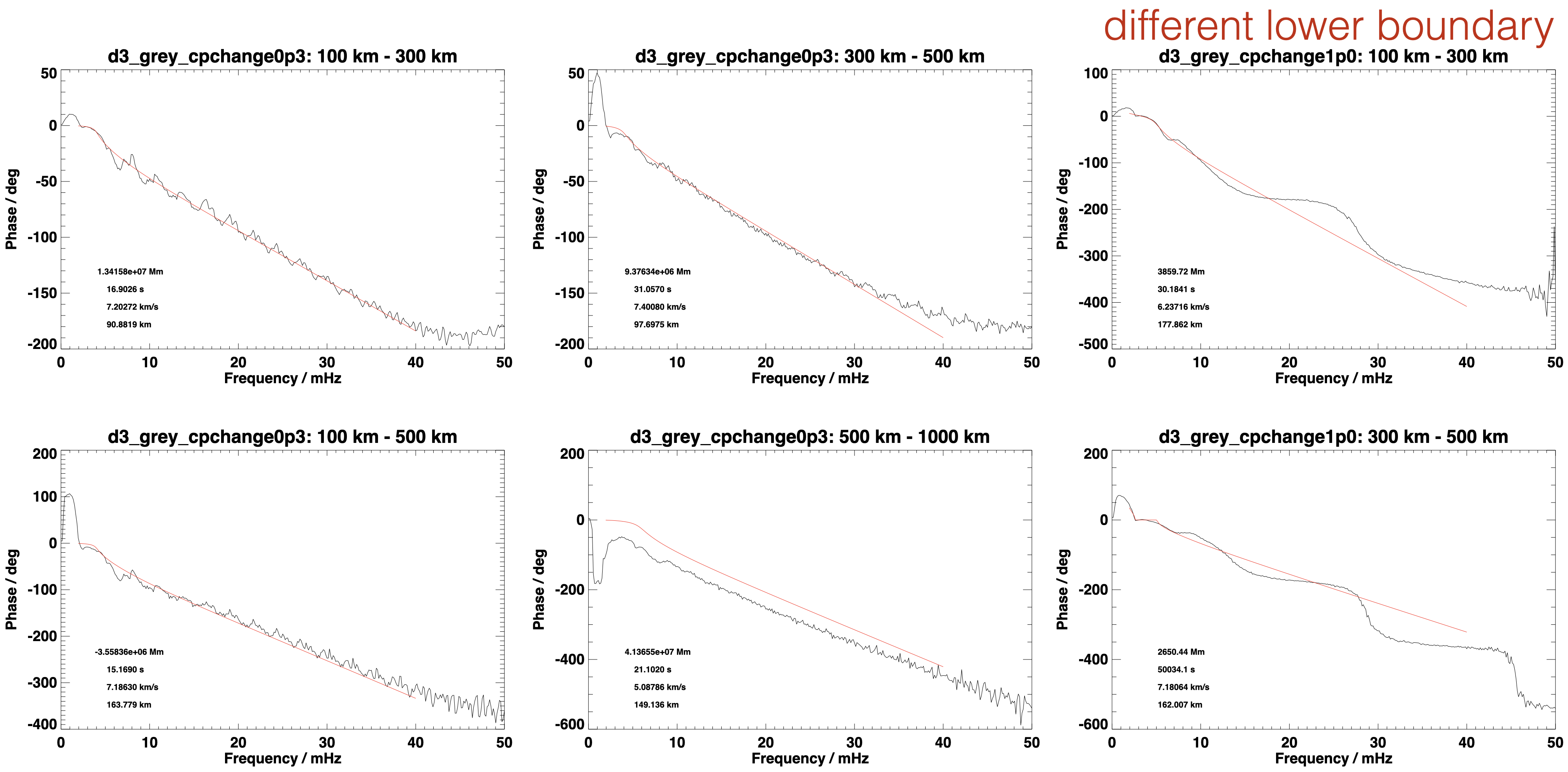}
\caption{Same as Fig.\ref{fig:1D-phase-Bifrost} for selected CO5BOLD models.}
\label{fig:1D-phase-Cobold}
\end{figure}

Fig.\,\ref{fig:1D-phase-Muram} shows similar diagrams from the three MURaM models (grey, non-grey, high and normal diffusivity). The measured phase difference in these models is strikingly different from those in the Bifrost models, with a monotonous linear decrease to the maximum frequency displayed here (50 mHz, for consistency with the results from the other models), with no indication of a "plateau" or phase jump. This linear trend can be followed up to the Nyquist frequency of these models ($\approx 200$\,mHz). The Souffrin fits (which were weighted with the measured power at the lower level) agree remarkably well with the spectra from the simulations. Interestingly, there is no significant differences visible between the three model runs, i.e. for the dispersion relation of the waves propagating in the MURaM models it does not seem to matter whether radiation was treated in grey or non-grey, and even the diffusivity does not have a measurable impact.

Fig.\,\ref{fig:1D-phase-Mancha-Muram} displays the phase difference spectra between the four levels in the MANCHA3D model and for comparison two spectra from the grey MURaM simulation. Interestingly, the Mancha simulations also indicate signs of a "wobble" in the lower atmosphere, although to a much lesser extent than the Bifrost simulations. In addition, the MANCHA3D model also reveals a full 180$^\circ$ phase jump near 30 mHz, similar to the 180$^\circ$ phase jump in the Bifrost spectrum between these two heights. No indication of such a phase jump is visible in the corresponding MURaM spectrum. 180$^\circ$ phase jumps usually are the signs of wave reflection and the formation of a node. There are no indications of such in the power spectra and their evolution with height, however (cf. Figs\,\ref{fig:power_spectra} and \ref{fig:sim_rms_vs_height}).

Finally, in Fig.\,\ref{fig:1D-phase-Cobold} we display similar phase difference spectra from the grey CO5BOLD model runs. The left and middle columns show spectra from the cp0p3 model. There, no signs of a "wobble", phase plateau or phase jump are visible. We do not understand the origin of the high-frequency ripple superimposed on the linear decrease in this model. The phase spectra displayed at the right are from the cp1p0 model, which uses a stiffer 
lower boundary. This model does not have the high frequency "ripples", but shows distinct, nearly horizontal phase plateaus (in particular in the 300 -- 500 km spectrum), separated by relatively sharp phase transitions. In a longer time series with higher frequency resolution, these might steepen to 180$^\circ$ phase jumps (cf. the apparent dependence of the sharpness of these phase jumps in the Bifrost spectra in Fig.\,\ref{fig:1D-phase-Bifrost}). Note also the significant discrepancy of the measured spectrum to the Souffrin model in the 500 -- 1000 km spectrum (also of the MANCHA3D model in Fig.\,\ref{fig:1D-phase-Mancha-Muram}) . None of the models seems to reflect the evanescent behavior of waves in the 3 to 5 mHz range, with consequences also for the higher frequency waves. We do not yet understand the cause of this, but suspect that non-linearities developing above $\sim$ 700 km might be a factor.

\subsection{Phase difference $\nu - z$ cuts}

\begin{figure}
\centering
\includegraphics[width=0.8\textwidth]{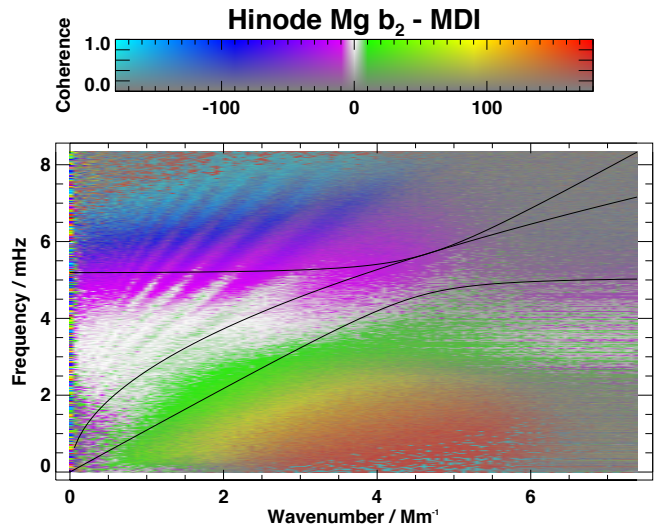}
\caption{$k$-$\omega$ 
phase difference spectrum between Doppler velocity oscillations measured with the Narrow-band Filter Imager (NFI) of Hinode's Solar Optical Telescope (SOT) in Mg b$_2$ and MDI in its high-resolution mode (Ni I 6768 {\AA}), overlaid by a diagnostic digram (thick black lines) and the location of the f-mode (thin black line).}
\label{fig:1Hinode_Mgb2-MDI}
\end{figure}

The 1-D phase difference spectra (as a function of frequency) presented and discussed in the previous section have several shortcomings, limiting their diagnostic value. First, the phase difference of acoustic-gravity waves is not only dependent on the temporal frequency, but also on the horizontal wavenumber $k_h$. This becomes evident in Fig.\ref{fig:1Hinode_Mgb2-MDI}, which shows a 
$k$-$\omega$ 
phase difference spectrum between Doppler velocity oscillations measured with the Narrow-band Filter Imager (NFI) of Hinode's Solar Optical Telescope (SOT, [25]) in Mg b$_2$ and SOHO/MDI [32] in its high-resolution mode (Ni I 6768 {\AA}). These data were obtained during a coordinated campaign on 2007/10/20. The observations covered a $109  \times 109$\,acrsec$^2$ square region near disk center ($256 \times 256$ pixels). The temporal resolution was 60\,s and the duration of the series 735\,min (12 hours 15 min). One can easily recognize the region of gravity waves with positive phase differences (green, yellow, red region at low frequencies and high wavenumbers below the Lamb line), the evanescent region of the 5 min p-modes with phase values close to 0$^\circ$ (whitish region above the f mode), and the region of propagating acoustic waves above the cut-off frequency (purple and blue colors). The saturation of the color is scaled by the coherence of the signal, i.e., regions of grey (mostly at high wave numbers) have low coherence. A 1-D (frequency) phase difference spectrum is the result of collapsing (averaging) the complex crosspower in $k_h$. As can be seen from Fig.\,\ref{fig:1Hinode_Mgb2-MDI}, acoustic waves are practically unaffected by this averaging over wavenumber, because the dispersion relation for waves above the cutoff frequency has very little dependence on the wavenumber (only at very high $k_h$), whereas in the lower frequency region ($\nu <  5$ mHz) the dispersion relation is strongly dependent on $k_h$.

The second and in this context more important reason why the 1-D phase difference spectra have limited diagnostic value is the fact that by looking at only selected heights, we cannot know what is happening in between these layers and where exactly some particular effect (like a wave node) might occur. To gain further insight, we therefore calculated two-dimensional  
$k$-$\omega$ 
phase difference spectra between successive levels, from bottom to top, for all simulations. The first step was to calculate 3-D Fourier transforms $\hat{v_i}(k_x,k_y,\nu)$ for all layers $v(x,y,z_i,t), i=1,\ldots,M$, with $M$ being the number of vertical layers in the models. From these, we calculated crosspower spectra between successive layers $z_i$ and $z_{i+1}$ for all M-1 layer pairs, and from those the phase difference after azimuthal integration in the 
$k_x$-$k_y$ 
plane (assuming horizontal isotropy). As a final step, we normalized the 
measured 
phase difference between the successive layers to a height difference of 100\,km, so that they are comparable in the various models. In the following we will show cuts 
at fixed horizontal 
wavenumbers 
$k_h = 2$\,Mm$^{-1}$ and $k_h = 4$\,Mm$^{-1}$ (cf. Fig.\,\ref{fig:1Hinode_Mgb2-MDI} for a location of these cuts 
in the diagnostic diagram) through the 3-D stacks of $k - \omega$ phase diagrams $\phi(\nu,k_h,z)$, which we will call $\nu-z$ phase diagrams.

\begin{figure}
\centering
\includegraphics[width=1.0\textwidth]{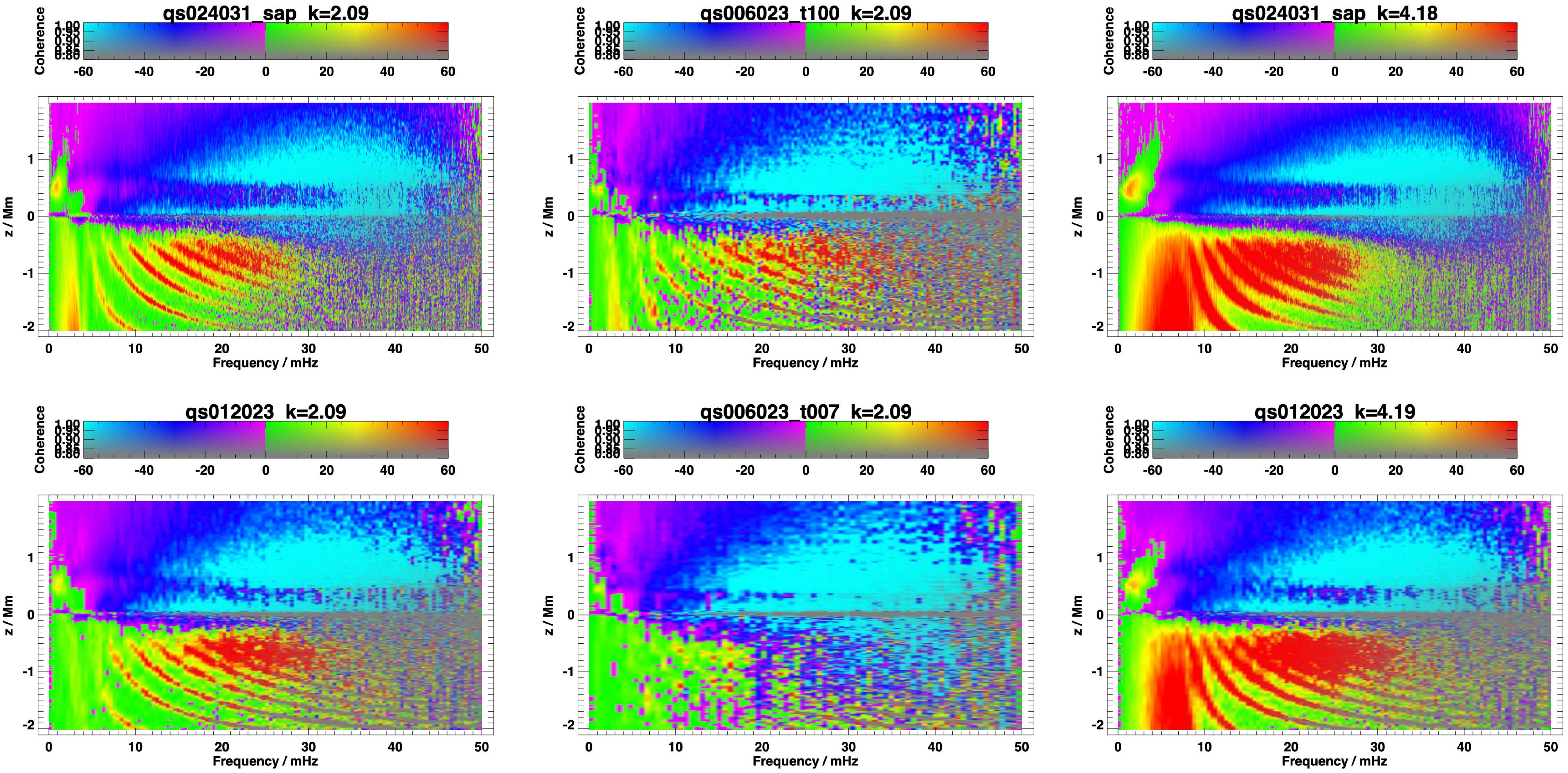}
\caption{$\nu$-$z$ 
phase diagrams (cuts 
at constant $k_h$ at approximately 2 and 4 Mm$^{-1}$ through the stacks of the layer-by-layer 2D 
$k$-$\omega$ 
phase difference diagrams) for various Bifrost models.}
\label{fig:nu-z-cut-Bifrost}
\end{figure}

In Fig.\,\ref{fig:nu-z-cut-Bifrost} we display $\nu-z$ phase diagrams at $k_h \approx 2$ and 4\,Mm$^{-1}$ of several Bifrost models. We chose these two $k_h$ values for the $\nu-z$ diagrams shown here, because power spectra of turbulent convection reach their maximum in this wavenumber range [7], and the cuts at these wavenumbers sample both gravity, evanescent, and acoustic waves (cf. Fig.\,\ref{fig:diagnostic-diagram} and Fig.\ref{fig:1Hinode_Mgb2-MDI}). Note that the field-of-view of most of the simulations we studied here is only in the range of $6\times6$\,Mm$^2$, resulting in a rather poor wavenumber resolution of only ~1\,Mm$^{-1}$. We could have  also shown cuts at 1, 3, or 5\,Mm$^{-1}$. These look very similar to the ones shown here, i.e. the conclusions would have been the same.

The $\nu-z$ phase diagrams displayed in Fig.\,\ref{fig:nu-z-cut-Bifrost} all show striking features, which are common to all models (even with the vastly different frequency resolution of the various models due to their different length): First, there is a dramatic difference between the subphotospheric layers (the convection zone) and the solar atmosphere above $z=0$\,km ($\tau_{5000} = 1$) layer. Except for the qs006023\_t007 model (which has a different lower boundary condition), the subphotospheric layers are dominated by "ridges" of constant, strongly positive phase values (red features in Fig.\,\ref{fig:nu-z-cut-Bifrost}) on a background of only slightly positive phase values (green areas). We cannot offer an explanation for this striking behavior yet, merely some speculative ideas: Assuming that the sound waves are created near the upper boundary of the convection zone, one would have expected a similar behavior as in the atmosphere, just with the opposite sign, i.e., downward propagating acoustic waves, showing up as positive phase differences, increasing with frequency (from green over yellow to red). The fact that we see phase "ridges" suggests an interference pattern, possibly caused by the reflection of the downward propagating sound waves at the lower boundary. These "ridges" may therefore be related to the pseudo-modes [32] observed in the atmosphere (see also Fig.\ref{fig:1Hinode_Mgb2-MDI}, where the modal structure of the p-modes can be followed beyond the acoustic cutoff frequency of about 5.5\,mHz). These high-frequency pseudo-modes are caused by the interference of waves coming directly from the source and those being reflected in the solar interior [32]. While acoustic waves with frequencies below the cutoff frequency are completely reflected by the surface layers and thus being trapped in the interior forming a pattern of standing waves, waves with frequencies above the cutoff frequency can escape into the solar atmosphere. Clearly, this finding requires further study, supported by focused modeling efforts.

In the atmosphere ($z > 0$\,km), we see in large parts the expected behaviour: green and red areas at the lowest frequencies in the gravity wave regime, followed by a transition into propagating acoustic waves with increasingly negative phases differences (purple and blue regions; also see Fig.\,\ref{fig:1Hinode_Mgb2-MDI} for comparison). There is one feature, though, which caught our attention and which sheds some light on the mysterious "wobbles" and 180$^\circ$ phase jumps in the 1-D Bifrost phase difference spectra. This is the "finger''
of increased phase speed (smaller than expected phase differences; purple and dark blue colors, rather than light blue) extending into the high frequency range between approximately 100 and 700 km height. This feature is most visible in the qs024031\_sap model (upper left and upper right in Fig.\,\ref{fig:nu-z-cut-Bifrost}), which is also the Bifrost model run with the largest length and hence best frequency resolution, but indications of this "finger" of anomalous phase speed are present in the other model runs as well. It is also interesting to see that in the chromosphere above $z \gtrsim 1300$\,km, the increase of the phase differences with frequency is reduced (darker blue rather than light blue), i.e. the phase speed is increased. Might this be related to the conundrum of the high phase speeds (vanishing phase differences) observed between Doppler oscillations of chromospheric lines [24, 34]?

\begin{figure}
\centering
\includegraphics[width=1.0\textwidth]{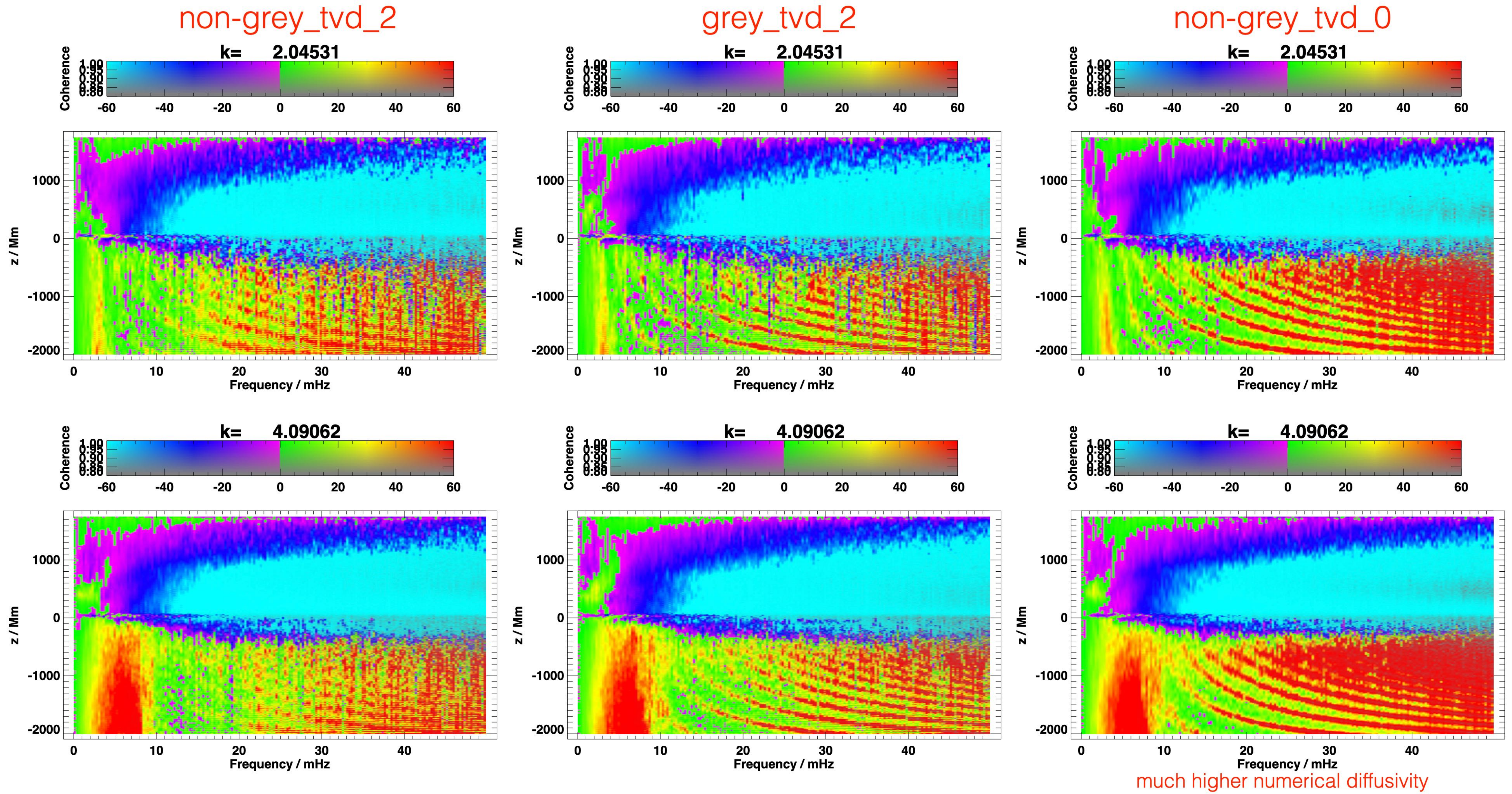}
\caption{Same as Fig.\,\ref{fig:nu-z-cut-Bifrost} for the three MURaM models.}
\label{fig:nu-z-cut-Muram}
\end{figure}

\begin{figure}
\centering
\includegraphics[width=1.0\textwidth]{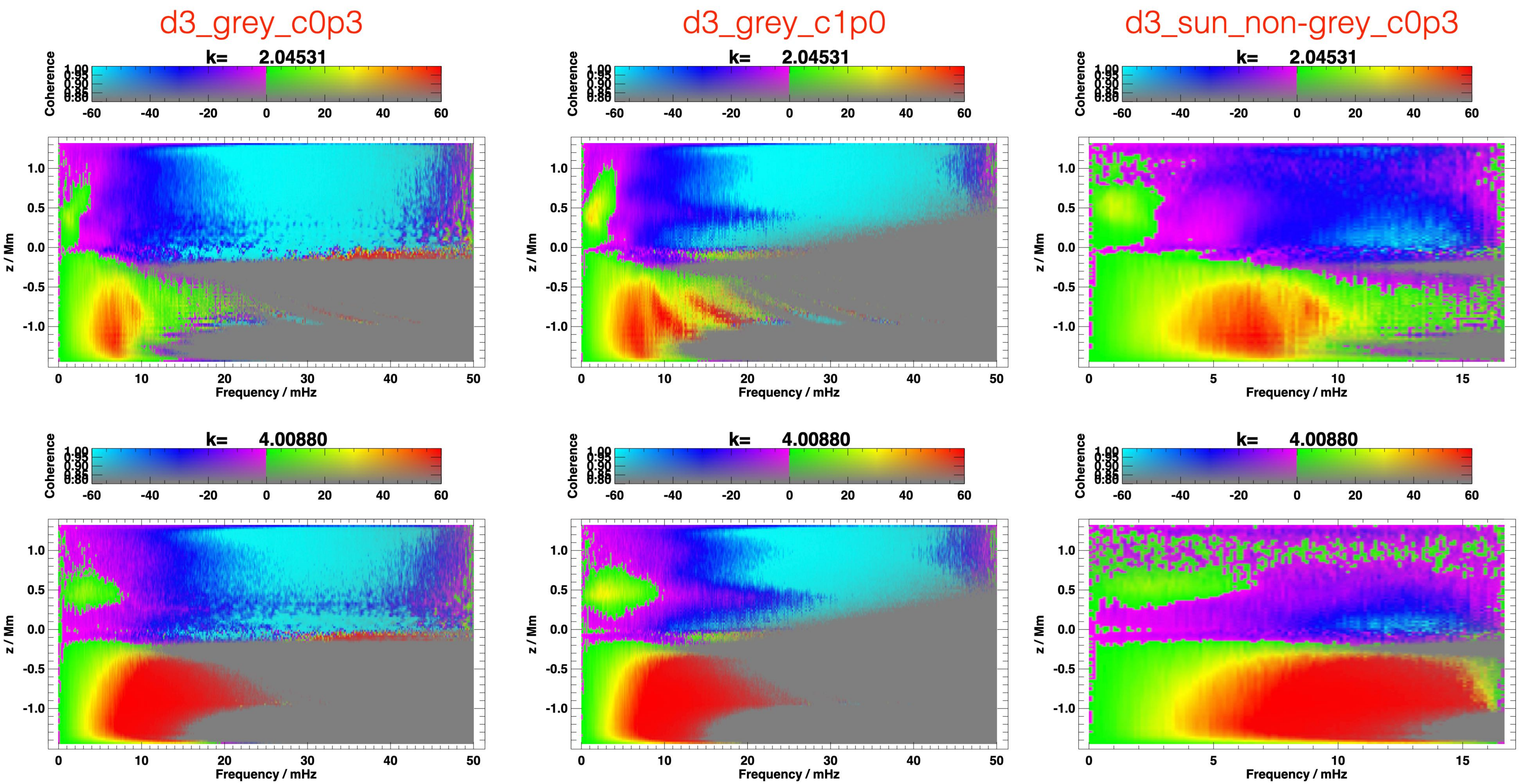}
\caption{Same as Fig.\,\ref{fig:nu-z-cut-Bifrost} for the CO5BOLD models.}
\label{fig:nu-z-cut-Cobold}
\end{figure}

\begin{figure}
\centering
\includegraphics[width=1.0\textwidth]{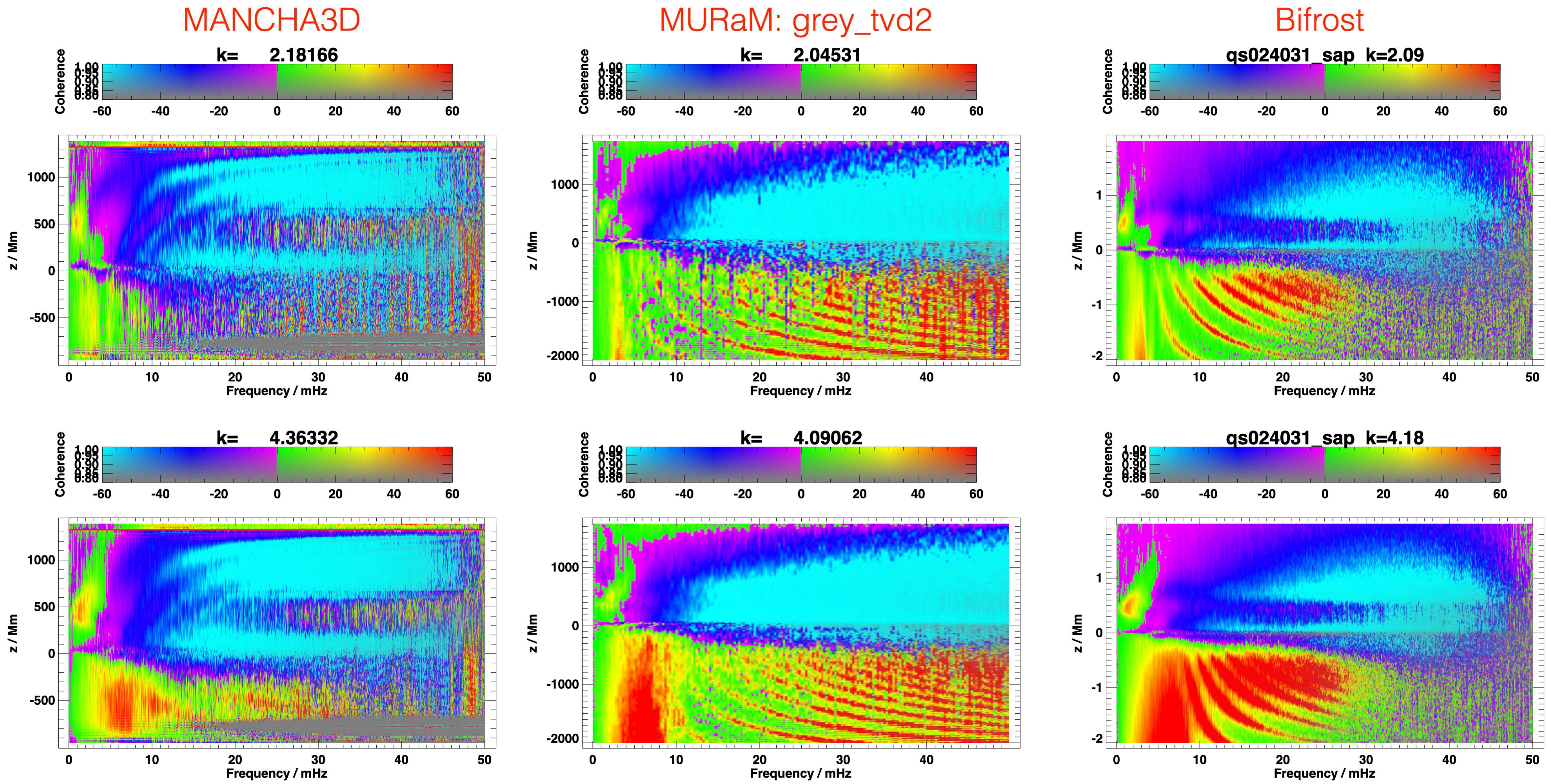}
\caption{Same as Fig.\,\ref{fig:nu-z-cut-Bifrost} for the MANCHA3D model and comparison to a MURaM and Bifrost model.}
\label{fig:nu-z-cut-Mancha}
\end{figure}

Fig.\,\ref{fig:nu-z-cut-Muram} shows the corresponding $\nu$-$z$ 
phase diagrams for the three MURaM models at $k_h \approx 2$ Mm$^{-1}$ (top 
row) and $k_h \approx 4$ Mm$^{-1}$ (bottom 
row). They reveal the same striking features as the Bifrost models: a clear separation in phase propagation behavior between the subphotospheric layers and atmosphere, with very prominent phase "ridges" in the convection zone. In the atmosphere, one can recognize the signature of gravity waves in the "photosphere" and the expected behavior of increasing phase lags with frequency in the acoustic range. There are no indications of the "finger" that is so prominent in the low to middle "photosphere" in the Bifrost simulations. Interestingly, the highest layers above 1000\,km also show smaller phase lags, i.e. higher phase speeds.

The $\nu$-$z$ 
phase diagrams of the CO5BOLD
models 
(Fig.\,\ref{fig:nu-z-cut-Cobold}) look quite different from the Bifrost and MURaM models. All the higher frequencies in the convection zone, where the other two models revealed the very striking phase ridges, have low coherence in this model (grey areas). Note the different frequency scale in the two non-grey simulations in the right column, which extends only to 16.6 mHz. The acoustic range shows the expected behavior, at least for the grey simulations. The c1p0 model (middle column) shows a similar "finger" of reduced phase differences (high phase speeds) between approximately 200 and 600\,km height. Interestingly, in distinction from the MURaM simulations, where there is practically no difference visible between the grey and non-grey case, for the CO5BOLD models the radiative treatment seems to make a significant difference (compare the 
first two columns with the rightmost one).

Finally,  in Fig.\,\ref{fig:nu-z-cut-Mancha} we compare the $\nu$-$z$ 
phase diagrams of the MANCHA3D model with the grey MURaM model and the qs024031\_sap Bifrost model. The differences are quite striking. Close to its lower boundary, the coherence in the MANCHA3D model is extremely low. Further, it does not show the prominent phase ridges in its convection zone that are so obvious in the MURaM and Bifrost models. While it also reveals signatures of gravity waves at low frequencies (most pronounced at around 500\,km height), it reveals conspicuous "ridges" in the acoustic wave regime between approximately 5 and 20 mHz, where one would expect a smooth gradient from purple over dark blue into light blue (cf. the MURaM results in the middle column of Fig.\,\ref{fig:nu-z-cut-Mancha}). Further, at high frequencies ($\nu > 20$\,mHz) there is a region of extremely noisy phase signal between about 200\,km and 600\,km height in the atmosphere. We recall that it is in this region where the 1D MANCHA3D phase difference spectra revealed a 180$^\circ$ phase jump. We cannot offer an explanation for the three high frequencies ridges or why the coherent phase propagation is lost in the MANCHA3D simulations between approximately 200 and 600\,km.

\section{Conclusion}
In an effort to benchmark the dynamics in simulations of the solar atmosphere, we have compared the wave propagation characteristics in various model runs produced with the Bifrost, CO5BOLD, MANCHA3D and MURaM codes. We have studied the height dependence of wave power in the various models, compared 1-D phase difference spectra between selected heights, and investigated the phase propagation of acoustic gravity waves from layer to layer, bottom to top, of the simulations. All models reveal the signatures of acoustic-gravity waves in their atmospheres, i.e., both gravity waves as well as acoustic waves are present in all models. The power distribution is relatively similar, but there are variances of up to two orders of magnitude at higher frequencies. It may be premature to use these models for the purpose of estimating the energy flux by acoustic or gravity waves without further detailed comparisons. While the phase difference spectra between selected heights overall show the expected behavior (certainly up to the frequencies which have been well observed), many models show deviations from the  linear decrease expected for propagating acoustic waves at high frequencies, including 180$^\circ$ phase jumps. There are regions in some models (typically in the "photosphere" between about 200 and 600 km height) where we detected significantly reduced phase differences (i.e., high phase speeds). These "fingers" are most prominent in 
the Bifrost models, 
but are also clearly visible in models of MANCHA3D and to some extent also in CO5BOLD models. What is the cause of these "fingers" of high phase speed? We do not know yet. 

Interestingly, the differences between model runs from a particular code using a different radiative transfer treatment (grey vs non-grey) or different lower boundary conditions are usually smaller than the differences between the various simulation codes. There is practically no difference in the wave propagation characteristics between the grey and non-grey MURaM model runs, whereas there are significant differences for the CO5BOLD models. 
What can we learn from the presence or absence of these differences?

There are indications of higher phase speeds (vanishing phase differences) in the "chromosphere" 
of all models. Might this lead to an explanation for the long-standing conundrum of the vanishing phase differences measured between chromospheric lines?

Particularly striking are the phase "ridges" in the 
$\nu$-$z$ 
phase diagrams in the convection zones of most models, in particular the Bifrost and MURaM models. We cannot offer a full explanation for these yet. Maybe they are formed by a similar mechanism as the pseudo-modes observed in the solar atmosphere. Clearly, these deserve further studies.

One of the shortcomings of the present study is the fact that the models from the various codes were in many aspects quite different. 
Some had 
magnetic fields, some were pure hydrodynamical; 
the fields of view, depths of the convection zone and vertical extent into the "chromospheres", 
spatial resolution, box size, duration, and diffusivities were all quite different. In a next step, to better understand the significant differences between models of the various codes (cf. Fig.\,\ref{fig:sim_rms_vs_height} and Fig.\,\ref{fig:nu-z-cut-Mancha}), we will attempt to repeat this study 
using the same (magneto-)hydrodynamic 
setup, same box size and resolution, same 
depth of the lower and upper boundaries, for both a low and high diffusivity case with grey and non-grey radiative transfer. If we still see differences that are larger than those we can attribute to different numerical diffusivities, we will have to dig deeper and worry about the detailed implementations of the physics, or ask ourselves if this approach of comparing wave propagation characteristics may have some issues as well. We will also conduct some linear wave propagation experiments in order to separate the complexities of non-linear dynamics from the wave propagation part and the ability of the codes to propagate different wave modes. Finally, for a quantitative comparison with observations, one needs to derive synthetic spectral lines and Doppler shifts from the models rather than using directly the velocities at a fixed geometrical height of the models.

\vskip6pt

\enlargethispage{20pt}


\dataccess{The SOHO/MDI data used in this study are available from the Joint Science Operations Center (JSOC) at Stanford University (\href{http://jsoc.stanford.edu}{http://jsoc.stanford.edu}) and the Hinode/SOT data from the Hinode Science Data Centre Europe at the University of Oslo, Norway (\href{http://sdc.uio.no/sdc/}{http://sdc.uio.no/sdc/}). The simulations this study is based on are too large to host on public repositories. However, parts of the data can be requested from the corresponding author, who will be happy to discuss ways to access the data.}

\aucontribute{BF conceived the study, carried out the analysis and drafted the manuscript. All others contributed their simulation models, participated in the discussion of the results and helped polish the manuscript.}

\competing{The authors declare that they have no competing interests.}


\ack{The authors wish to acknowledge scientific discussions with the Waves in the Lower Solar Atmosphere (WaLSA; www.walsa.team) team and the Royal Society through the award of funding to host the Theo Murphy Discussion Meeting "High resolution wave dynamics in the lower solar atmosphere" (grant Hooke18b/SCTM). This research was supported by the Research Council of Norway through its Centres of Excellence scheme, project number 262622, and through grants of computing time from the Programme for Supercomputing. This material is based upon work supported by the National Center for
Atmospheric Research, which is a major facility sponsored by the National Science Foundation under Cooperative Agreement No. 1852977. EK was supported by the Spanish Ministry of Science through the grant PGC2018- 095832-B-I00, and by the European Research Council throught the grant ERC-2017-CoG771310-PI2FA. Computing resources of the Spanish Supercomputing Network (LaPalma and MareNostrum supercomputers) are thankfully acknowledged. Hinode is a Japanese mission developed and launched by ISAS/JAXA, collaborating with NAOJ as a domestic partner, NASA and STFC (UK) as international partners. Scientific operation of the Hinode mission is conducted by the Hinode science team organized at ISAS/JAXA. This team mainly consists of scientists from institutes in the partner countries. Support for the post-launch operation is provided by JAXA and NAOJ(Japan), STFC (U.K.), NASA, ESA, and NSC (Norway). SOHO is a mission of international cooperation between ESA and NASA.}



\end{document}